Meteoroid Ablation within the Jovian Atmosphere Implications on the Oxygen Delivery to the Gas Giant's Atmosphere

C A Mehta[1,2*], D M Orlov[2,3], T Feng[4], R W James[1], and E G Kostadinova[2]

[1]The United States Coast Guard Academy. New London, CT 06320

[2]Auburn University. Auburn. AL, 36849

[3]University of California, San Diego. San Diego, CA 92093

[4]University of South Florida. Tampa, FL 33620

**\*Correspondence**

C A Mehta, *The United States Coast Guard Academy, New London, CT* 06320.
Christopher.A.Mehta@uscga.edu

The detection of water molecules within the atmosphere of Jupiter, first by the Galileo Atmospheric Probe, and later by the Juno spacecraft, has given rise to the question of whether those molecules are sourced endogenously or exogenously. One hypothesis is that the ablation of meteoroids deposited the necessary oxygen into the planet's atmosphere, which was subsequently used in chemical processes to form water. This paper aims to evaluate this hypothesis by simulating the ablation of carbonaceous objects entering the planet's atmosphere to determine the possible rates of oxygen delivery and the most likely altitude for such delivery within the Jovian atmosphere. We estimate that carbonaceous meteoroids have the potential to deliver ~1.4 x $10^7$ kg/m² of oxygen over a billion years. We further estimate that most of the ablation is expected to occur in the stratosphere, or 450-400 km above the region of 1 bar of atmospheric pressure. In comparison, Interplanetary Dust Particles (IDPs) are estimated to deliver roughly ~$10^2$ kg/m² of oxygen over the same period.

1. **Introduction**

The ablation of objects entering the atmosphere of Jupiter has been of interest due to the potential of delivering amino acids, sugars, and other hydrocarbons to the planet through meteoritic sources (proposed by Sagan et al., 1967). Meteoroid ablation in the planet's atmosphere has also been suggested as a possible mechanism for the formation of ions within the ionosphere, which can lead to the further formation of compounds (Kim et al., 2001). Past studies showing the presence of water in the Jovian atmosphere concluded that the needed oxygen was supplied by exogenous sources (Feuchtgruber et al., 1997; Moses and Poppe, 2017). This reasoning is based on the observation that water condenses at lower depths of the planet's troposphere, while endogenous reactions rely on oxygen upwelling. From that depth, the element upwelling was shown to be an ineffective source of the amount of oxygen needed to produce water (Feuchtgruber et al., 1997; Moses and Poppe, 2017). Although photochemical reactions of oxygen ions are thought to be an additional source for the production of water in the planet's atmosphere, these processes still do not fully account for the amount of oxygen needed to promote water production on the planet, per Poppe, 2016. Here, we use numerical modeling to study the ablation of carbonaceous meteoroids as a mechanism for the delivery of oxygen to various altitudes within the Jovian atmosphere and discuss whether this process could potentially account for the detected amount of water in Jupiter's upper atmosphere. What follows is a brief overview of the structure and composition of Jupiter's atmosphere with an emphasis on water in the atmosphere, along with a summary of meteoroid size distributions used in our numerical modeling. Information on the simulation methods and ablation model used for the study is provided in Section 2. Section 3 includes the simulation results along with a discussion pertaining to the transport of oxygen from meteoritic sources to Jupiter. A summary of conclusions is provided in Section 4.

## 1.1 Oxygen Photochemical Reactions in Jupiter's Atmosphere

The Jovian atmosphere can be broken down into the troposphere (the innermost layer), stratosphere, thermosphere, and exosphere (the outermost layer). The surface of the planet can be defined as the region in which the atmospheric pressure is 1-bar (Sieff et al., 1998). The atmospheric temperature decreases from the bottom of the exosphere to roughly 320 km above the surface (the stratosphere-thermosphere boundary) and maintains approximately constant value until the top of the tropopause, ~50 km above the surface (Seiff et al., 1998). Below the tropopause, the planet is expected to be saturated with clouds composed of ammonia and water (Moses and Poppe, 2017; Seiff et al., 1998). Jupiter's atmosphere is primarily composed of hydrogen and helium, with the presence of hydrogen-rich compounds in the atmosphere consistent with a bulk composition similar to that of the Sun (Ingersoll 1976 and Irwin 1999). In addition to hydrogen and helium, the atmosphere of Jupiter contains trace amounts of other compounds, including methane, ammonia, water, and hydrogen sulfide, with these species appearing as the most abundant after hydrogen and helium (Ingersoll 1976 and Irwin 1999). Figure 1 shows estimated water and oxygen percentages with respect to different atmospheric layers. Estimates were obtained from Rensen et al., 2023 and Moses and Poppe, 2017. Water within Jupiter can be synthesized from reactions occurring within the atmosphere with the help of exogenous sources. In the event of an exogenous object entry, as the object ablates, various elements are released into the atmosphere of the planet. These elements can subsequently undergo either thermochemical reactions within the tail of the impactor or photochemical reactions, both can alter the planet's atmospheric composition (Connerney and Waite, 1984; Giles et al., 2021; Majeed and McConnell, 1991; Moses and Poppe, 2017; Ollivier et al., 2000). The oxygen ions being supplied to the planet via the ablation of meteoroids can undergo the following chain of reactions to produce water (Strobel, 2005):

$$O^+ + H_2 \rightarrow OH^+ + H \qquad (1)$$

$$OH^+ + H_2 \rightarrow H_2O^+ + H \qquad (2)$$

$$H_2O^+ + CH_4^+ \rightarrow H_3O^+ + CH_3 \qquad (3)$$

$$H_3O^+ + e \rightarrow H_2O + H \text{ (or alternatively } OH + H_2) \qquad (4)$$

Poppe, 2016 coupled numerical modeling with data from the Juno mission to study the flux of dust particles from cometary and Kuiper Belt Objects. This was done to gain insight into the potential sources of elements that can be used in photochemical reactions within the atmospheres of gas giants. Findings from their study show that photochemistry of material from interplanetary dust particles (IDPs) alone does not fully account for the delivery of oxygen to Jupiter (Poppe, 2016). This motivates the need to study additional mechanisms, including objects much larger than dust, to gain insight into the delivery of oxygen and the related formation of water in the Jovian atmosphere. The study by Moses and Poppe (2017) estimated the delivery of oxygen to Jupiter from Interplanetary Dust Particles (IDPs) to be around $10^6$ to $10^7$ oxygen atoms per square centimeter per second (or ~$10^1$-$10^2$ kg/m² of oxygen over a billion years). In contrast, data from the Juno spacecraft indicated that the actual amount of oxygen present in Jupiter's atmosphere was significantly higher than what could be accounted for by IDPs alone (Szalay et al., 2024; Cavalié et al., 2023; Fonte et al., 2023)). Additionally, while lightning detected on Jupiter by Voyager and other spacecraft implied the presence of water, an accurate estimate of the amount of water deep within Jupiter's atmosphere remained elusive (Ingersoll 1976 and Rensen et al. 2023). This discrepancy suggests that there must be additional sources or mechanisms contributing to the oxygen levels observed in Jupiter's atmosphere, such as ablation of larger exogenous objects. Exploring these additional mechanisms is necessary to fully understand the formation of water and the overall atmospheric composition of Jupiter. Here, we focus on carbonaceous meteorites and their potential to transfer oxygen to the Jovian atmosphere.

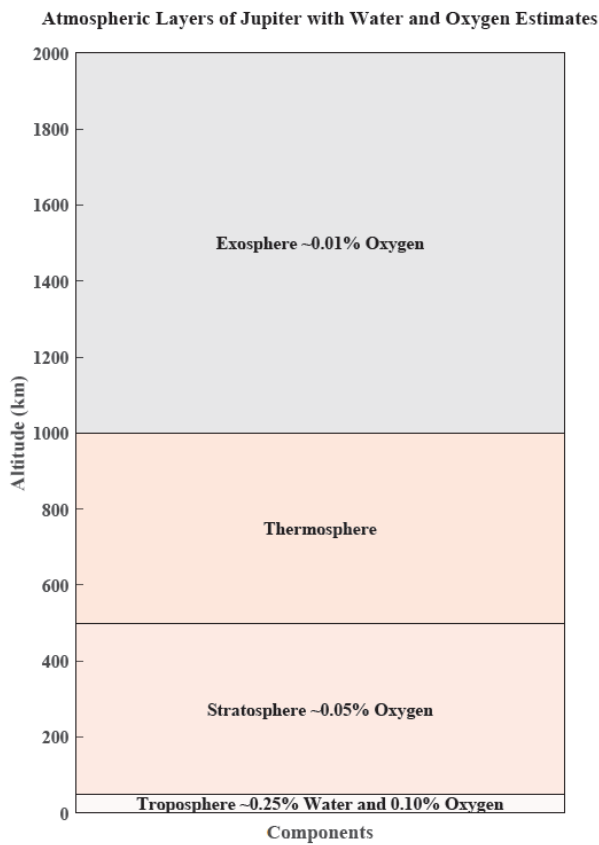

*Figure 1. Estimated oxygen and water percentage layers See Rensen et al. 2023 and Moses and Poppe, 2017 for in-depth estimates.*

1.3 Distribution of Jupiter Impact Events

The size distribution of asteroids in our Solar System has been investigated using a variety of methods, including the study of meteoritic dust clouds, examination of crater sizes, numerical modeling, and using radar systems (Nininger 1952; Huss 1991; Turcotte et al. 2003; Hawkes et al., 2002; Mathews et al., 2002; Zolensky et al., 2006; Pasek and Lauretta 2008; Badyukov and Dudorov 2013; Włodarczyk and Leliwa-Kopystyński 2014; Vinnikov et al., 2016; Ryabova 2017 Schult et al., 2018; and Rubincam 2018). Based on these studies, it is presumed that asteroid distribution follows a power-law, i.e., a distribution for which the number count (n) of objects of mass (m) is related linearly in log-log coordinate space, see equation (5). However, understanding the size distribution of smaller bodies (meteoroids) in our Solar System is challenging as such objects are more difficult to observe telescopically (Harris and D'Abramo, 2015). To this end, meteorites found on Earth may help constrain the power-law behavior of meteorite size distributions. If meteorites follow a power law (Newman, 2004 and He et al., 2014), as their parent bodies appear to do, then these relationships can be used to understand the mass distributions of objects entering not only Earth but other planets within our Solar System (James et al., 2018 and Tria et al., 2018). Per Bland and Artemieva, 2006, this relationship takes the form:

$$\log \log n = \alpha \log \log m + b \,, \qquad (5)$$

where $n$ is the number of samples with mass $m$ or greater, $\alpha$ is the power-law coefficient, and $b$ is the y-intercept (set by the number $n$ with $m = 1$ kg). Assuming that the log-log relationship found for carbonaceous chondrites on Earth holds true to Jupiter, it has been estimated that the Jovian planet is subjected to $5 \times 10^6$ impacts per year of objects with initial masses in the range 250 kg – 5,000 kg and $3 \times 10^5$ impacts per year of objects with initial mass greater than 5,000 kg (per Cook and Duxbury, 1981). More recently, the rate of impacts from data presented by (Hueso et al., 2018) has estimated that Jupiter is subjected to $2.6 \times 10^3$ – $1.7 \times 10^4$ impacts per year for objects in the range 250 kg – 5,000 kg and $0.16 \times 10^3$ – $1 \times 10^3$ impacts per year for objects greater than 5,000 kg. See (Giles et al., 2021) for an in-depth calculation and comparison for Jupiter's meteoritic flux. In this study, we assume that Jupiter is subjected to $3 \times 10^4$ impacts per year and use a power law distribution to cover the range of impactor masses.

Due to their abundance in our Solar System and their location primarily outside the asteroid belt, we focus on the ablation of carbonaceous chondrites as they enter Jupiter and investigate their possible role in transporting oxygen to the planet. First, to identify the atmospheric location where most of the ablation is expected to occur for a range of initial masses, we model the ablation of objects ranging from $10^3$ to $10^8$ kilograms for identical entry conditions. Next, assuming a power-law distribution of meteoroid masses, we employ a Monte Carlo Simulation to model $3 \times 10^4$ meteorites entering the planet for different entry conditions. This allows us to gain an understanding of how much material can be deposited into the atmosphere of the planet (above 1-bar of atmospheric pressure) over a one-year period. Finally, using this data, we calculate the amount of oxygen that could be transferred to the atmosphere of Jupiter, and we identify the region where we would expect to see an abundance of the element.

2. Methods

Meteoroid ablation occurs when the astromaterial enters an atmosphere, and the aerodynamic pressure of the surrounding atmosphere surpasses the material strength of the object. This generates heat from an atmospheric entry that breaks up the object, causing it to lose mass (Hughes, 1992; Bland et al., 1996; Bland and Artemieva, 2006). If, on impact, the kinetic energy exceeds the heat of vaporization of the object, the object starts vaporizing. When a meteoroid ablates, the forward velocity of the ablated material is equivalent to the velocity of the meteor. The ablated particles, along with the energy created by the ablation process, are then transported downstream of the ablating object, and the energy involved in the ablation process is transferred to heat the surrounding atmosphere (Zinn et al. 2004). The ablation model used in this study was adapted from Bland and Artemieva (2006), Pasek and Lauretta (2008), and Mehta et al. (2018). Considering various physical and atmospheric properties, we simulate the effect ablation has on a meteoroid transiting through the atmosphere. The model solves two equations that account for the change in velocity and change of mass with respect to time (Bland and Artemieva 2006):

$$\frac{dv}{dt} = -C_d \frac{\rho_g A v^2}{m} + g\sin(\theta) \quad (6)$$

$$\frac{dm}{dt} = -A \frac{C_h \rho_g v^3}{2Q}, \quad (7)$$

where $v$ is the initial velocity of the meteoroid, $m$ is the mass of the meteoroid, $A$ is the cross-sectional area of the meteoroid, $g$ is the acceleration due to gravity on Jupiter, $\theta$ is the angle of entry into the atmosphere, $C_d$ is the drag coefficient, $C_h$ is the heat transfer coefficient, and $Q$ is the heat of ablation. Together, $C_h/2C_dQ$ forms the ratio used to obtain the ablation coefficient. An ablation coefficient value of 0.042 s² km$^{-2}$ was used as it is the estimated average value for carbonaceous meteoroids per Ceplecha et al., 1998. Lastly, $\rho_g$ is the density of the Jovian atmosphere at a given height (defined by the planet's scale height or e-folding distance), calculated using:

$$\rho_g = D_{surf} * e^{-\frac{h}{z}}, \quad (8)$$

where $D_{surf}$ is the surface density, $h$ is the height above 1-bar, and $Z$ is Jupiter's scale height.

Using these methods, we first simulate carbonaceous chondrites with mass ranging from $10^3$ to $10^8$ kilograms entering the atmosphere of Jupiter using fixed values of the entry angle and velocity based on estimated averages. This is used as a control to identify the regions within the Jovian atmosphere where objects lose most of their mass due to ablation. Thereafter, we simulate 30,000 meteoroids (slightly over the yearly flux presented in Giles et al., 2021 to account for smaller objects) entering the planet's atmosphere with initial mass, velocity, and angle of entry from a Monte Carlo approximation to estimate Jupiter's mass flux of oxygen. Initial meteoroid masses were randomly selected from a power-law distribution (per Bland and Artemieva, 2006). The smallest mass was set to 1 kg, and the largest mass was selected by the Monte Carlo simulation (9 x $10^5$ kg). The entry velocity was normally distributed around an average initial velocity of 51,000 m/s (assuming most objects enter Jupiter at the planet's escape velocity) with a standard deviation of 6000 m/s. The angle of entry was selected between 1-90° with a mean of 30° and standard deviation of 10° (see Lui et al. 2019). For simplicity, we only consider spherical meteoroids.

### 3. Results and Discussion

3.1 Carbonaceous Meteoroid Ablation

To understand where most of the ablation takes place within Jupiter's atmosphere, we simulate objects ranging from $10^3$ to $10^8$ kg (large meteoroids), entering at 51 km/s and at 30 degrees. Calculations show that smaller objects (~1,000 kg) tend to vaporize in the upper stratosphere of the planet. As objects increase in initial mass, they penetrate further through the atmosphere prior to vaporization. Nonetheless, the region of the atmosphere that yields the most ablation was calculated to be the stratosphere of the planet, 450-400 km above the region of 1 bar of atmospheric pressure (see Figure 2). We observe that 90-99% of the ablation occurs in this region for most large meteoroids (~$10^7$ kg), withstanding the ablation process longer and, thus, vaporizing at lower altitudes. Therefore, we expect that any ablated oxygen that is ionized would be available for chemical reactions within this region of the planet's atmosphere. The comparatively steep slopes associated with Figure 2 are a clear representation of the drastic influence Jupiter's atmosphere has on the ablation process. Typically, higher atmospheric densities increase frictional heating, resulting in steeper mass loss slopes (Rogers et al., 2005). In addition, thermal ablation typically results in steep mass loss slopes, especially for meteoroids entering at high velocities (as intense heating causes rapid evaporation of the meteoroid's material) (Brykina and Egorova, 2023). Another factor that reflects the steep slope values observed is the material properties associated with carbonaceous meteoroids (as they typically require more heat to ablate) (Ceplecha et al., 1998; Bland and Artemieva, 2006; and Popova, 2004).

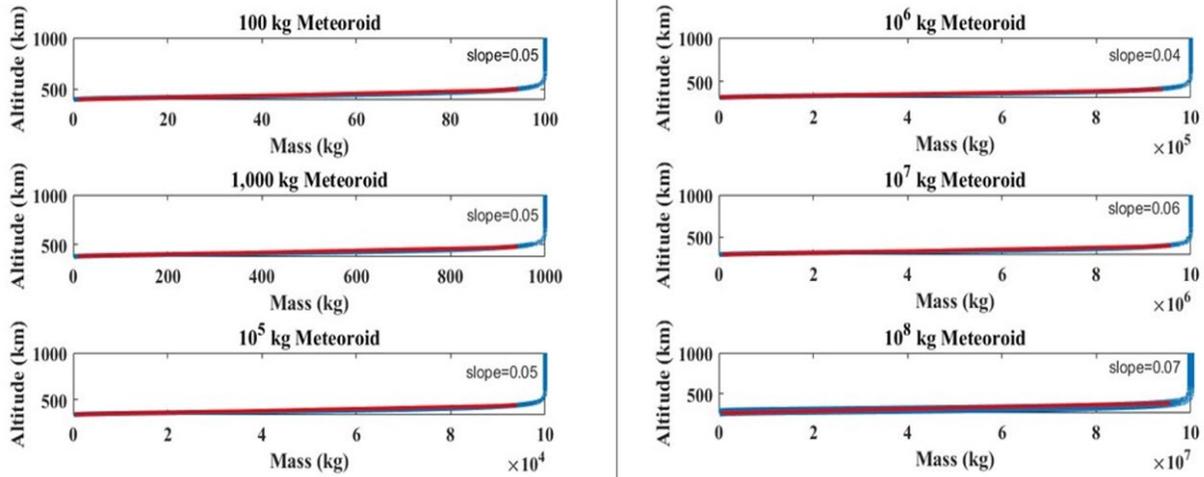

*Figure 2. Ablation of small to large spherical objects.*

Table 1. Parameters used for the ablation model to determine the region of *Jupiter's* atmosphere where most of the ablation is expected to take place for meteoroids.

| Scale height | Starting altitude | Drag Coefficient | Angle of Entry | Gravitational acceleration | Entry velocity | Meteoroid density | Ablation coefficient |
|---|---|---|---|---|---|---|---|
| 27 km | 1,000 km | 0.47 | 30° | 0.02479 km/s$^{-2}$ | 51 km/s | 3.5 g/cc | 0.042 s$^2$km$^{-2}$ |

## 3.2 Monte Carlo Approximation and Oxygen Flux Calculation

To calculate an estimated flux of oxygen deposited into Jupiter's atmosphere, we simulated 30,000 carbonaceous meteoroids entering the planet with initial mass, initial velocity, and angle of entry drawn randomly through a Monte Carlo approximation to assess the possible amount of oxygen that could be transported to the planet. The smallest mass was set to 1 kg, and the largest mass selected by the Monet Carlo simulation was 9 x 10$^5$ kg, with smaller mass entries occurring more frequently (Figure 3).

Table 2. Values used in Monte Carlo Approximation. SD stands for standard deviation.

| Smallest Mass | Largest Mass | Initial Velocity | Angle of Entry |
|---|---|---|---|
| 1 kg | 9 x 10$^5$ kg | 51 km/s (SD: 6 km/s) | 1-90° (SD: 30°) |

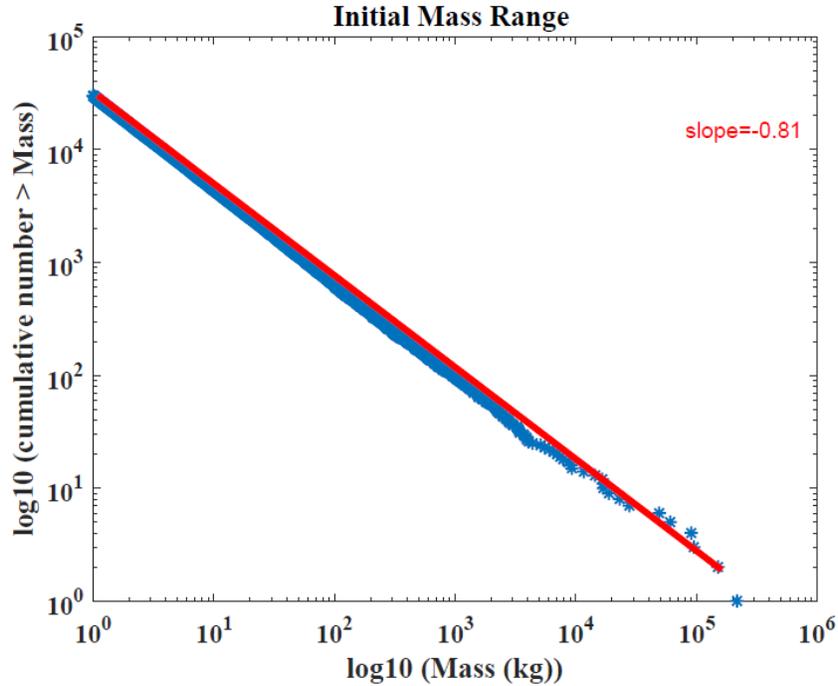

Figure 3. Power-law distribution of initial masses used in this study.

Figures 4 and 5 summarize the effect of the angle of entry and initial velocity on the final height (height of vaporization/total ablation) of meteoroids entering Jupiter. We see that the angle of entry affects the location where objects vaporize within the planet. Objects entering at lower angles tend to withstand the ablation process for a longer time than those objects entering at a larger angle. Meteoroids that enter the atmosphere at angles between ~15 to 45 degrees vaporize between 450-400 km. This, of course, is also related to the entry velocity, such that objects entering the atmosphere at higher velocities are subjected to vaporization faster than objects entering at lower velocities (Figure 6).

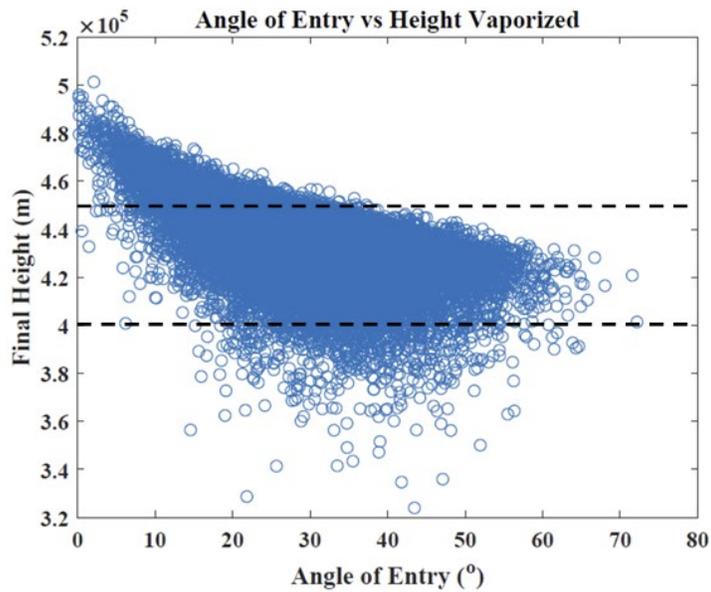

*Figure 4. Angle of entry as a function of height of vaporization. dashed line region within the Jovian atmosphere in where most ablation occur.*

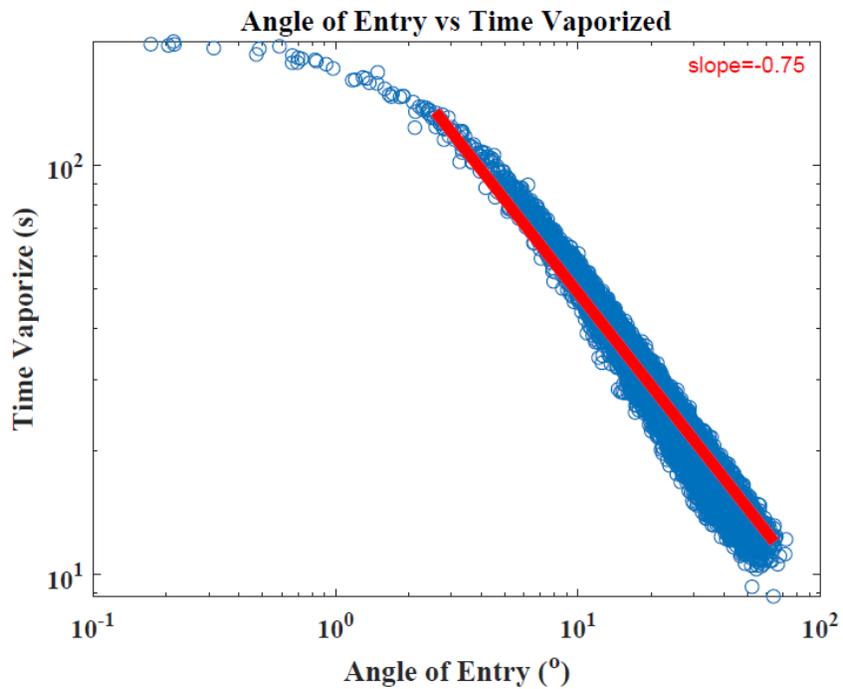

*Figure 5. Angle of entry with respect to time of vaporization. Objects entering the Jovian atmosphere at a shallow entry are generally able to withstand ablation better than those entering at high angle of entry.*

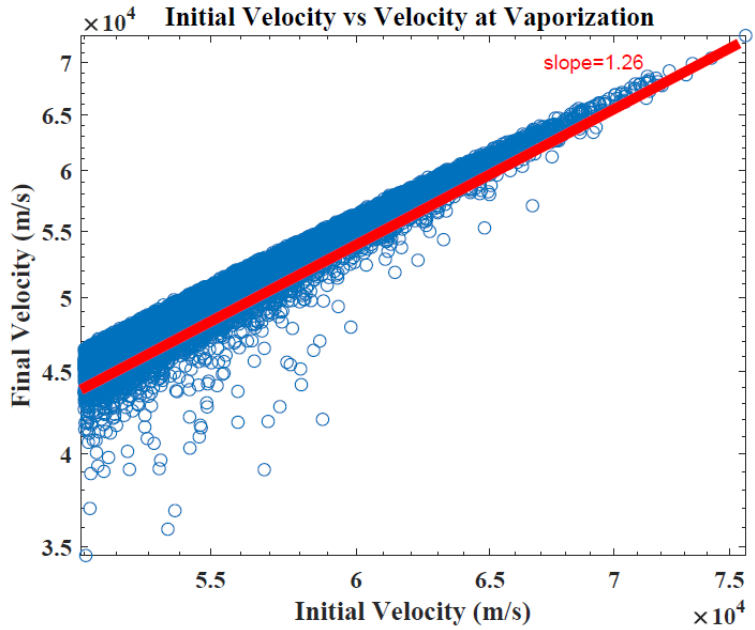

*Figure 6. Initial vs Velocity when vaporized.*

Assuming there are ~3 x 10³ meteoroid impacts per year and integrating the used power law distribution of masses, we estimate that the total mass delivered to the planet through meteoroid ablation is ~2.24 x 10⁴ kg/yr. We expect that 90-99% of that mass is delivered in the planet's stratosphere, or 450-500 km above the region of 1-atmospheric bar. Furthermore, if we assume that carbonaceous meteoroids have a maximum $\partial^{18}O$ abundance of 19.30% per Greenwood et al., 2023, we can calculate the amount of oxygen delivered to the upper atmosphere of Jupiter using the following equation from Pasek and Lauretta, 2008

$$F_x = M_F x_E k. \quad (9)$$

Here $F_x$ is the flux of the element of interest, $M_F$ is the yearly meteoritic mass flux, $x_E$ is the abundance of elements of interest, and k is the factor determining potentially prebiotic varieties of the element. We assume that oxygen present in carbonaceous meteoroids would be in the form of $\partial^{18}O$ due to the high abundance of 19.30% per Greenwood et al., 2023, with 20% weighted percent of water (Howard et al., 2015; Trigo-Rodríguez et al., 2019). Thereafter, the steady-state concentration of oxygen in the planet's atmosphere is calculated using the following equation from Pasek and Lauretta, 2008:

$$|X| = \frac{F_x t_{1/2}}{ln(2)}, \quad (10)$$

where $|X|$ is the concentration of oxygen and $t_{1/2}$ is the half-life of the element of interest ($\partial^{18}O$ being stable). Using these equations, we estimate that the planet receives an oxygen flux of roughly 14.05 x 10⁻³ kg/m², spread across the surface area of the planet over one year or 1.405 x 10⁷ kg/m² of oxygen over a billion years (compared to oxygen delivered by smaller, interplanetary dust particles transferring ~10¹⁻10² kg/m² of oxygen over a billion years per Moses and Poppe 2017).

## 3.3 Shape Dependent Meteoroid Ablation and Product Formation Through Meteoritic Ionization/Atmospheric Shock

Different shapes ablate at different rates (Figure 7). As a result, the amount of oxygen that could potentially be deposited into the Jovian atmosphere strongly varies not only as a function of mass but also as a function of initial shape. Furthermore, carbonaceous chondrites micrometeoroids are thought to rarely be spherical in shape (Kurat et al., 1994). To this end, it is important to investigate how non-spherical meteoroids ablate when entering Jupiter. When modeling three $10^5$ kg meteoroids (spherical, conical, and conical with blunt side down) with an entry velocity of 51 km/s and angle of entry of 30 degrees, we see that conical meteoroids withstand the ablation process better than spherical meteoroids. Therefore, if conical objects withstand the ablation better, this leads to a potential decrease the material deposited in the Jovian atmosphere at higher altitudes. Alternatively, it may mean that the oxygen may be delivered at lower altitudes.

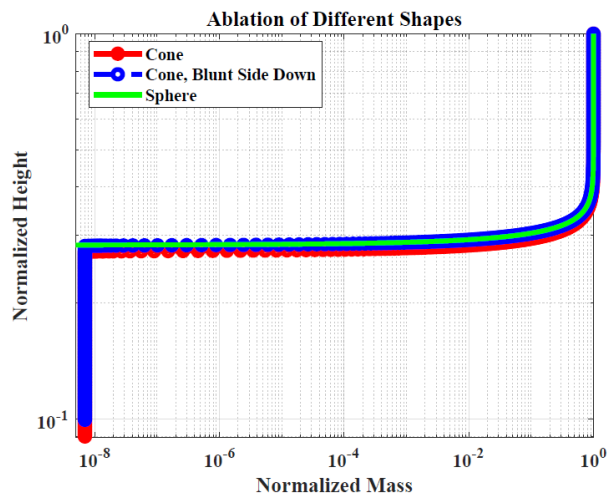

*Figure 7. Ablation of Sphere, Cone, and Blunt Side Down Cone. Note, spherical objects tend to ablate faster than conical objects.*

Another avenue for creating chemical compounds in Jupiter is through plasma chemistry in the tails of meteoroids, shocking the atmosphere as they enter the planet (Campbell and Brunger, 2018, 2019; Moses, 1992; Moses and Poppe, 2017). With the amount of elements deposited into the atmosphere, specifically in the tail of the ablating object, the heat generated by the atmospheric entry process is enough to produce the formation of products. Therefore, one potential hypothesis is that the total concentration of water could be accounted for, not only by meteoroids depositing ablated material in the form of heavy elements, but also by meteoroids themselves shocking the atmosphere and the ablated materials interacting with the heat and atmospheric gasses from the atmosphere. For example, meteoritic ionization of water in a hydrogen-dominant atmosphere could produce a hydroxyl molecule (Campbell and Brunger, 2019). If this is the case, then this combination of the endogenous and exogenous sources could further account for the high amounts of oxygen-bearing molecules present in Jupiter's atmosphere. Therefore, future studies should aim to augment our understanding of how organics and inorganics can form within a plasma environment. Recent experiments at the DIII-D tokamak were conducted by our team with the goal of studying plasma chemistry in heating conditions reminiscent to those expected for tails of ablating meteoroids. The experimental design was very similar to the one described in (Orlov et al., 2021.). Preliminary results suggest that these heating conditions are favorable for the formation of ammonia. Detailed analysis of these experiments will be presented in a separate publication.

## 4. Conclusion

In this study, we employed numerical modeling to understand how carbonaceous meteoroids entering Jupiter ablate and how this affects the amount of oxygen that can potentially be deposited into the atmosphere of the planet. To do so, we simulated 30,000 meteoroids entering the atmosphere with the initial conditions of the object distribution drawn using a Monte Carlo approximation based on data collected by past studies of gas giant impactors. Although all the 30,000 objects simulated in this study eventually vaporized, objects entering the planet favored low-angle entries, with the time to vaporization slightly decreasing with respect to the low angle of entries. Small, medium, and large objects entering the planet's escape velocity tend to vaporize in the upper stratosphere of the planet, with objects entering at lower velocities (although very rare) penetrating through the atmosphere further than those entering at higher velocities. Using this data, we calculated an estimated flux, spread across the planet, to be $1.405 \times 10^7$ kg/m² of oxygen over a billion years that can be utilized in various atmospheric chemical processes. This, combined with the total flux of oxygen from Interplanetary Dust Particles ($10^2$ kg/m² of oxygen over the same period), could account for the formation of oxygen-bearing molecules we see in the upper atmosphere (above ~450 km) of Jupiter, which merits more investigation, specifically tailored towards modeling of water formation within the planet's atmosphere.

Although our model provides a good first-order approximation of the ablation of objects as they enter Jupiter, this research can benefit from a model that considers other processes, such as plasma chemistry in the tail of the ablating meteoroids. Furthermore, we consider only spherical meteoroids, while carbonaceous chondrites are seldom uniformly spherical (Kurat et al., 1994). Future studies should aim to augment the current model to account for such processes, which will give us a better understanding of the transport of matter from exogenous sources and the formation of atmospheric compounds. Future works should also be supplemented by laboratory studies of materials ablation utilizing plasma conditions similar to atmospheric entry conditions. Specifically, understanding how the formation of compounds (inorganic and organic) could potentially be formed by meteoritic atmospheric shock/ionization on not only Jupiter but also Earth.

**Acknowledgments**

The authors would like to thank the United States Department of Energy for providing funding for this project under DE-SC0023375, DE-SC0022554, DE-SC0021338, and DE-FC02-04ER54698. Furthermore, we would like to thank Dr. Matthew Pasek, Dr. Lorin Matthews, Dr. Truell Hyde, Dr. Dennis Bodewits, Dr. Augusto Carballido, Dr. Babak Shotorban, Dr. Bryant Wyatt, Dr. Youssef Moulane, and Dr. Mohi Saki for their valuable insight and comments on this project.